\documentclass[aps,eqsecnum,nofootinbib,superscriptaddress]{revtex4}
\usepackage{amsmath}
\usepackage{amsfonts}
\usepackage{amssymb}
\usepackage[dvips]{graphicx}  
\usepackage{amsmath,amsfonts}
\usepackage{color}  
\def\be{\begin{eqnarray}}
\def\ee{\end{eqnarray}}
\def\beq{\begin{equation}}
\def\eeq{\end{equation}}

\def\({\left (}
\def\){\right )}
\def\S{{\cal S}}

\newtheorem{theorem}{Theorem}[section]
\newtheorem{lemma}[theorem]{Lemma}
\newtheorem{proposition}[theorem]{Proposition}
\newtheorem{corollary}[theorem]{Corollary}
\newtheorem{definition}{Definition}[section]
\newtheorem{remark}[theorem]{Remark}

\newcommand{\qed}{\nobreak \ifvmode \relax \else
      \ifdim\lastskip<1.5em \hskip-\lastskip
      \hskip1.5em plus0em minus0.5em \fi \nobreak
      \vrule height0.75em width0.5em depth0.25em\fi}
\newcommand{\qcd}{\begin{flushright} $\Box$ \end{flushright}}
\begin{document}

\title{
On the geodesic incompleteness of spacetimes containing marginally outer trapped surfaces   
}

\author{I.P. Costa e Silva}
\email{ivanpcs@mtm.ufsc.br}
\affiliation{Department of Mathematics,\\ 
Universidade Federal de Santa Catarina \\88.040-900 Florian\'{o}polis-SC, Brasil}

\date{\today}

\begin{abstract}
In a recent paper \cite{EGP}, Eichmair, Galloway and Pollack have proved a Gannon-Lee-type singularity theorem based on the existence of marginally outer trapped surfaces (MOTS) on noncompact initial data sets for globally hyperbolic spacetimes. This result requires that the MOTS be {\em generic} in a suitable sense. In the same spirit, this author has proven some variants of that result \cite{me2} which hold for weaker causal conditions on spacetime, but which concern (generic) marginally trapped surfaces (MTS) rather than MOTS, i.e., most of the results in \cite{me2} need a condition on the convergence of the ingoing family of normal null geodesics as well. However, much of the more recent literature has focused on MOTS rather than MTS as quasi-local substitutes for the description of black holes, as they are arguably more natural and easier to handle in a number of situations. It is therefore pertinent to ask to what extent one can deduce the existence of singularities in the presence of MOTS alone. In this note, we address this issue and show that singularities still arise in the presence of generic MOTS under weaker causal conditions (specifically, for causally simple spacetimes). Moreover, provided we assume that the MOTS is the boundary of a compact spatial region, a Penrose-Hawking-type singularity theorem can be established for chronological spacetimes containing generic MOTS.     
              
\end{abstract}
%
%
\maketitle

\section{Introduction}\label{sec:intro}

The existence of a closed trapped surface in spacetime is a well-known, fundamental criterion for gravitational collapse. This concept was introduced by Roger Penrose \cite{P} as an abstract indicator that the region containing it, say, describing some stellar core, had attained a ``point of no return'' of matter concentration, and the formation of a black hole would presumably follow, at least if the collapse would not deviate too much from spherical symmetry. A mathematical justification for the appropriateness  of this criterion stems from the fact that, under physically reasonable additional assumptions, the existence of such surfaces implies nonspacelike geodesic incompleteness of the spacetime manifold, which is the geometrical counterpart of gravitational collapse.       

However, the existence of black holes {\em cannot} be directly deduced from either the existence of closed trapped or from the existence of singularities, unless one assumes that some form of the Cosmic Censorship conjecture holds (see, e.g., Ch. 12 of \cite{wald} for a nice discussion). Part of the difficulty involved here lies in the {\em global} nature of the black hole and event horizon concepts, which makes it difficult to establish their existence from local considerations. This global character also makes the standard mathematical notion of black hole notoriously ill-suited to handle more realistic analyses of astrophysical black hole candidates, especially in their dynamical aspects, as well as in numerical studies. Accordingly, the theoretical focus in recent years has shifted to a systematic study of {\em quasi-local} notions of horizons. Among these are {\em trapped}, {\em isolated} or {\em dynamical} horizons (see, e.g., \cite{AK} for extensive discussions and references), whose chief defining feature is to be foliated by {\em marginally trapped surfaces} (MTS), understood as codimension two closed submanifolds with nonpositive convergence of the two normal families of null geodesics. 

Another quasi-local notion especially suited for the study of dynamical black holes at the level of initial data sets is the {\em apparent horizon}, i.e., the boundary (in an underlying spatial slice of spacetime) of the region containing closed trapped surfaces. Apparent horizons have become increasingly important in recent times, not only because of the current emphasis on quasi-local notions, but also because they were found to have a number of mathematically interesting properties. Indeed, when smooth they are known to be {\em marginally outer trapped surfaces} (MOTS), codimension two closed submanifolds in which the {\em outgoing} family of normal geodesics is constrained and required to have zero convergence (see, e.g. chapter 12 of \cite{wald} and Ref. \cite{AM}), but no restriction is made on the ingoing family. Now, both MOTS and MTS are natural initial data surrogates for event horizons, but MOTS are in some respects easier to study. For example, proving the existence of a MTS directly is more involved than that of a MOTS, because it entails control on the sign of certain quantities, which in general is a difficult problem \footnote{I thank Marc Mars, José Senovilla and Michael Eichmair for their comments on this point.}. MOTS, on the other hand, have been extensively studied in the Mathematical Physics literature, and some existence results for them have been proven \cite{AEM,AMMS,AM,eichmair1,eichmair2}. Apart from their appearance in Relativity, they are particularly interesting to geometers because they have a number of properties similar to minimal surfaces, and can be thought of as natural generalizations of the latter \cite{mots2}. Moreover, they play an important role in the Schoen-Yau's proof of the Positive Mass theorem \cite{SY1,SY2}. 

There is another, more fundamental reason to consider MOTS in connection with gravitational collapse. For while the existence of closed trapped surfaces is the mathematical criterion {\em par excellence} of the onset of gravitational collapse, a direct {\em proof} that they naturally appear for large enough concentrations of mass is not forthcoming, except of course in specific models. To the best of this author's knowledge, the closest approach to such a general proof is the celebrated Schoen-Yau's result in Ref. \cite{SY3}, which however proves the existence of MOTS rather than that of closed trapped surfaces. The question then naturally arises as to the conditions under which one can infer that the existence of MOTS implies gravitational collapse. In a recent paper \cite{EGP}, Eichmair, Galloway and Pollack investigate the existence of MOTS in initial data sets, and they find that MOTS naturally form in asymptotically flat initial data sets with non-trivial topology; in addition, they prove that if the existence of such a MOTS $\Sigma$ contained in a noncompact Cauchy hypersurface in spacetime is granted, and this MOTS is moreover ``generic'' in the sense that future and past inextendible null geodesics normal to $\Sigma$ have non-zero tidal acceleration somewhere along them, then null geodesic incompleteness follows. (The null convergence condition is also assumed.) This result can therefore be understood as a Gannon-Lee-type \cite{gannon1,gannon2,lee,me} singularity theorem. Clearly, some sort of genericity condition is needed to infer incompleteness. For instance, by performing simple isometric identifications in Minkowski spacetime one can check that there are even globally hyperbolic spacetimes which are geodesically complete while having compact marginally (outer) trapped surfaces. (See also \cite{AMMS}, where a singularity theorem in globally hyperbolic spacetimes containing MOTS is proven, with alternative generic conditions on the MOTS.)   

It is the purpose of this note to investigate such singularity theorems in the presence of MOTS with causality conditions weaker than global hyperbolicity valid for all spacetime dimensions $n$ greater than 2. In a recent paper \cite{me2}, some such results were indeed obtained, but for MTS rather than MOTS. Specifically, one can prove a Hawking-Penrose-type theorem (after the main theorem in Ref. \cite{penrose}) for generic MTS which, in causal terms, assumes only the non-existence of closed timelike curves \cite{me2}. In this paper we consider generic MOTS (to be defined more precisely below), so no assumption will be made on the ingoing family of normal geodesics. In this context, we argue, by means of simple examples, that in general one should not hope to weaken too much the causal requirements unless one imposes extra specific conditions either on the spacetime or on the MOTS (other the the usual assumptions in the main singularity theorems). We then prove a version of the singularity theorem in \cite{EGP} valid for {\em causally simple} spacetimes. Causal simplicity is a condition only slightly weaker than global hyperbolicity, but which holds in important examples as Kerr-Newman (extended maximally except for the causality-violating region) or anti-de Sitter spacetimes. But if the MOTS bounds a compact spatial region (as will be the case, for instance, if the region of the spatial slice containing it is diffeomorphic to some $\mathbb{R}^n$), then one can go all the way to prove a Hawking-Penrose-type theorem for generic MOTS which, in causal terms, assumes only the non-existence of closed timelike curves. In the case $n=4$, this result together with those in Ref. \cite{SY3} go a long way towards establishing on a {\em model-independent}, mathematically rigorous footing the basic heuristic expectation that large concentrations of mass do lead to gravitational collapse.

The rest of the paper is organized as follows. We first give some preliminary definitions in Section \ref{preliminaries} to set the conventions, general assumptions and notation, and two propositions (Propositions \ref{mainproposition1} and \ref{mainproposition2}) which will be key ingredients in the proof of our main results. The proofs of the main results, Theorems \ref{mainsingularity1} and \ref{mainsingularity2}, together with some simple consequences, are deferred to Section \ref{mainstuff}.

\section{Preliminaries: Basic Definitions \& Results}\label{preliminaries}

 In what follows, we fix a {\em spacetime}, i.e, an $n$-dimensional, second-countable, connected, Hausdorff, smooth (i.e., $C^{\infty}$) time-oriented Lorentz (signature $(-,+, \ldots,+)$) manifold $M$ endowed with a smooth metric tensor $g$ and with $n \geq 3$. We assume that the reader is familiar with the basic definitions and results of global Lorentzian geometry and causal theory of spacetimes as found in the core references \cite{HE,wald,oneill,BE}, and in particular with the standard singularity theorems. We denote by $d: M \times M \rightarrow [0,+\infty]$ the (lower semicontinuous) Lorentzian distance function on $(M,g)$, and by $L_g(\gamma)$ the Lorentzian arc-length of a causal \footnote{The adjectives `causal' and `nonspacelike' as applied to curves (or curve segments) are used interchangeably.} curve segment $\gamma:[a,b] \rightarrow M$. All submanifolds of $M$ are embedded unless otherwise specified, and their topology is the induced topology. Finally, we follow the convention that causal vectors are always nonzero.
 
 We start by recalling a few standard definitions and results (cf. chapter 8 of Ref. \cite{BE} for examples and further discussion). Our intention here is merely to establish additional notation and terminology which will be used later on.  

\begin{definition}
\label{raysnlines}
A {\em future-directed timelike [resp. null] geodesic ray} in $(M,g)$ is a future-directed, future-inextendible timelike [resp. null] geodesic $\gamma : [0,b) \rightarrow M$ ($b\leq +\infty $) for which $d(\gamma(s),\gamma(t)) = L_g(\gamma |_{[s,t]})$ for all $s,t \in [0,b)$ with $s \leq t$.  In either case (i.e., timelike or null), $\gamma$ is also called a {\em future-directed nonspacelike geodesic ray}. 
\end{definition}

A {\em past-directed} timelike [resp. null, nonspacelike] geodesic ray can be analogously defined. In what follows, unless explicitly stated otherwise, we shall always consider {\em future-directed} causal curves, and we henceforth drop explicit references to the causal orientation. According to the above definition, a nonspacelike geodesic ray or is characterized by the fact that it {\em maximizes the Lorentzian arc-length between any two of its points}. If $\gamma$ is a {\em null} ray or line, the condition that $d(\gamma(s),\gamma(t)) = L_g(\gamma |_{[s,t]})$ for all $s,t$ in the domain of $\gamma$ with $s \leq t$ is actually equivalent to the requirement that the image of $\gamma$ be {\em achronal} in $(M,g)$, i.e., no two of its points can be connected by a timelike curve segment.

Recall that an achronal set $A \subseteq M$ is {\em future-trapped} if $E^{+}(A) = J^{+}(A)\setminus I^{+}(A)$ is compact [a {\em past-trapped} set is defined time-dually]. 

Let $\S \subset M$ be a smooth, connected, spacelike, partial Cauchy hypersurface \footnote {Recall that a {\em partial Cauchy hypersurface} is by definition an acausal edgeless subset of a spacetime, which means in particular that it is a closed topological (i.e. $C^0$) hypersurface \cite{oneill}. In this paper, however, we always deal with {\em smooth} hypersurfaces.} (submanifold of codimension one), and a smooth, with each connected component of $\Sigma$ a compact (without boundary), spacelike submanifold of codimension two (loosely called {\em surface} in what follows) $\Sigma \subset \S$. We assume that $\Sigma$ is two-sided in $\S$. This means, in particular, that there are unique unit spacelike vector fields $N_{\pm}$ on $\Sigma$ normal to $\Sigma$ in $\S$.    

For simplicity we shall assume throughout this paper that $\Sigma$ is connected and {\em separates} $\S$, i.e., that $\S \setminus \Sigma$ is not connected. (This assumption is not essential for the results in this paper, however, for one can always focus on one connected component and consider a covering of $M$ in which this holds for that connected component of $\Sigma$. Now, in dealing with geodesic incompleteness one may as well work in covering manifolds.) Thus, $\S \setminus \Sigma$ is a disjoint union $\S_{+}\dot{\cup} \S_{-}$ of open submanifolds of $\S$ having $\Sigma$ as a common boundary. We shall loosely call $\S_{+}$ [resp. $\S_{-}$] the {\em outside} [resp. {\em inside}] of $\Sigma$ in $\S$. (In most interesting examples there is a natural choice for these.) The normal vector fields $N_{\pm}$ on $\Sigma$ are then chosen so that $N_{+}$ [resp. $N_{-}$] is outward-pointing [resp. inward-pointing], i.e., points into $\S_{+}$ [resp. $\S_{-}$].     

Let $U$ be the unique timelike, future-directed, unit normal vector field on $\S$. Then $K_{\pm} := U|_{\Sigma}+N_{\pm}$ are future-directed null vector fields on $\Sigma$ normal to $\Sigma$ in $M$. The {\em expansion scalars} of $\Sigma$ in $M$ are the smooth functions $\theta_{\pm}: \Sigma \rightarrow \mathbb{R}$ given by 
\begin{equation}
\label{expansion}
\theta_{\pm}(p) = - \langle H_p, K_{\pm}(p)\rangle _{p},
\end{equation}
for each $p \in \Sigma$, where $H_p$ denotes the mean curvature vector of $\Sigma$ in $M$ at $p$ \cite{oneill}, and we denote $g$ as $\langle \, , \, \rangle$ here and hereafter, if there is no risk of confusion. 

Henceforth, whenever we refer to a {\em surface $\Sigma$ contained in a partial Cauchy hypersurface $\S$}, all the above conventions will be understood. 

Physically, the expansion scalars measure the divergence of light rays emanating from $\Sigma$. If $\Sigma$ is a round sphere in a Euclidean slice of Minkowski spacetime, with the obvious choices of inside and outside, we have $\theta_{+}>0$ and $\theta_{-} < 0$. One also expects this to be the case if $\Sigma$ is a ``large"  sphere in an asymptotically flat spacetime. But in a region of strong gravity one expects instead that we have both $\theta_{\pm} < 0$, in which case $\Sigma$ is a {\em closed (future) trapped surface}. $\Sigma$ is a {\em marginally outer trapped surface} (MOTS) is $\theta_{+} = 0$. It is well known (see, e.g, p. 310 of \cite{wald}) that either MTS or MOTS remain inside the black hole region, provided they are in a strongly asymptotically predictable spacetime in which the null convergence condition holds. 

We shall consider the following additional notion: 
 
\begin{definition}
\label{genericMOTS}
 A MOTS $\Sigma$ is {\em generic} (in $(M,g)$) if any future-directed, future-inextendible, or past-directed, past-inextendible null geodesic $\eta:[0,a) \rightarrow M$ ($0<a\leq +\infty$) starting at $\Sigma$ and normal to $\Sigma$ at $\eta(0)$ satisfies the generic condition, i.e., at some point $p$, and for some vector $v$ normal to $\eta'(p)$, $\langle v,R(v,\eta')\eta' \rangle \neq 0$.      
\end{definition}

The generic condition formulated above is a mild constraint which ensures (see, e.g., Prop. 2.11 in \cite{BE}) that there is non-zero tidal acceleration somewhere along each null geodesic $\eta$ normal to $\Sigma$. This will occur, for example (see Prop. 2.12 in \cite{BE}), if $Ric(\eta',\eta') \neq 0$ somewhere along $\eta$, which in turn, if the Einstein field equations hold for $(M,g)$, will likely happen whenever $\eta$ crosses (or is part of) some matter-energy cluster of positive density.  

\begin{definition}
Let $\eta:[0,a) \rightarrow M$ be a null geodesic normal to $\Sigma$ at $\eta(0)$. $\eta$ is said to be {\em outward-pointing} [resp. {\em inward-pointing}] if $\eta'(0)$ is parallel to $K_{+}(\eta(0))$ [resp. $K_{-}(\eta(0))$]. It is a {\em $\Sigma$-ray} (conf. Definition 14.4 in Ref. \cite{BE}), i.e. the length of any segment starting at $\Sigma$ up to any point along the ray realizes the Lorentzian distance from $\Sigma$ to that point. In particular, there exists no timelike curve from $\Sigma$ to the ray and no focal points to $\Sigma$ along the ray.
\end{definition}

The key ingredients in the proofs of the main results in the next Section are the following two propositions. Proposition \ref{mainproposition1} will allow us to slightly generalize the singularity theorem in \cite{EGP} by isolating some key elements in the proof in a systematic manner, while Proposition \ref{mainproposition2} will be the basis for a Hawking-Penrose-type singularity theorem for certain MOTS.  

In Proposition \ref{mainproposition1}, we shall consider causally simple spacetimes. Recall that $(M,g)$ is {\em causally simple} if it is causal (i.e., has no closed causal curves) and $J^{\pm}(p)$ are closed sets for every point $p \in M$. It is not difficult to check that in that case, $J^{\pm}(K)$ are also closed for every $K \subset M$ compact. A detailed discussion of such spacetimes with some results and further references can be found, e.g., in Section 3 of Ref. \cite{me}. Of course, every globally hyperbolic spacetime is causally simple.    
 
In Ref. \cite{me} the notion of a {\em piercing} was also introduced, and we shall find it necessary to assume that one such exists in the Proposition \ref{mainproposition1} below. We recall the definition here. We say that a smooth future-directed timelike vector field $X:M \rightarrow TM$ is a {\em piercing} of $\S$ (or {\em pierces} $\S$) if every maximally extended integral curve of $X$ intersects $\S$ exactly once. In physical terms, one may think of the integral curves of a piercing as worldlines of members of a family of observers who ``witness"  the ``whole universe at a certain instant of common time" described by $\Sigma$. This interpretation will hopefully convince the reader that it is a rather harmless assumption from a physical standpoint. Moreover, it is not difficult to check that a piercing does exist for suitable partial Cauchy hypersurfaces in basic solutions like Minkowski, Kerr-Newman and FRW spacetimes.  

Of course, a piercing of $\Sigma$ may not exist for general spacetimes. On the other hand, if $(M,g)$ is globally hyperbolic and $\S$ is a Cauchy hypersurface, then {\em every} smooth future-directed timelike vector field in $M$ pierces $\S$. {\em However, the existence of a piercing for $\S$ is strictly weaker than the requirement that $\S$ be Cauchy}. For example, consider 4-dimensional anti-de Sitter spacetime, taken to be $\mathbb{R}^4$ with the metric given by the line element (see, e.g., \cite{HE}, pg. 131)
\begin{equation}
\label{AdS}
ds^2 = -\cosh ^2rdt^2 + dr^2 + \sinh ^2 r \left(d\theta ^2 + \sin ^2 \theta d \phi ^2 \right),
\end{equation}                
where the coordinate ranges are $-\infty < t < \infty$, $r >0$, $0< \theta <\pi$, and $0 < \phi < 2\pi$. This spacetime is not globally hyperbolic, but each hypersurface $t = \mbox{const.}$, although not Cauchy, is pierced by the vector field $\frac{\partial}{\partial t}$. 

\begin{proposition}
\label{mainproposition1}
Let $\Sigma \subset M$ be a surface contained in a partial Cauchy hypersurface $\S$. Suppose that the following holds.
\begin{itemize}
\item[i)] $(M,g)$ is causally simple. 
\item[ii)] $\overline{\S}_{+}$ is noncompact and $\S$ admits a piercing $X$. 
\end{itemize}
Then there exists an {\em outward-pointing}, future-directed $\Sigma$-ray starting at $\Sigma$ and normal to $\Sigma$. 
\end{proposition} 

{\em Proof.} The idea for the first part of the proof presented here is due to Gannon \cite{gannon2} (see also Theorem 7.1 in \cite{AMMS}), adapted to our more general setting. 

Note that because $(M,g)$ is causally simple, $E^{+}(\Sigma) = \partial I^{+}(\Sigma)$. Given $p \in E^{+}(\Sigma)\setminus \Sigma$, let $\eta:[0,1] \rightarrow M$ be a future-directed null curve with $\eta(1) = p$ and $\eta(0) \in \Sigma$. $\eta$ must of course be a segment of a null generator of $\partial I^{+}(\Sigma)$, and it must be normal to $\Sigma$ at $\eta(0)$. Therefore $\eta'(0)$ is either parallel to $K_{+}(\eta(0))$, or to $K_{-}(\eta(0))$. Hence, we denote by  ${\cal H}_{+}$ the set of points of $E^{+}(\Sigma)\setminus \Sigma$ for which the former case occurs, and by ${\cal H}_{-}$ the set of points in $E^{+}(\Sigma)\setminus \Sigma$ for which the latter case occurs. Thus $E^{+}(\Sigma)\setminus \Sigma = {\cal H}_{+} \cup {\cal H}_{-}$. 

We claim that ${\cal H}_{+} \cap {\cal H}_{-} = \emptyset $. Suppose not, and let $p \in {\cal H}_{+} \cap {\cal H}_{-}$. Let $\eta_{\pm}:[0,1] \rightarrow M$ be future-directed null curves with $\eta_{\pm}(1) = p$ and $\eta_{\pm}(0) \in \Sigma$, such that $\eta_{+}$ [resp. $\eta_{-}$] is outward-pointing [resp. inward-pointing].    Let $\eta: [0,1] \rightarrow M$ be given by  
\[
\eta(t) = \left\{\begin{array}{cc} \eta_{-}(2t),& \mbox{if $0 \leq t \leq 1/2$} \\ \eta_{+}(2-2t), & \mbox{if $1/2 \leq t \leq 1$} \end{array} \right. .
\]
Now, using the piercing $X$, we can define by a standard argument (see,e.g., proposition 31, ch. 14 of \cite{oneill}) a continuous open map $\rho_X:M \rightarrow \S$ leaving $\S$ pointwise fixed (i.e. a retraction) by ``sliding'' points of $M$ along the maximal integral curves of $X$ into $\S$.  Consider the continuous curve $\eta_{X}:[0,1] \rightarrow \S$ given by $\eta_{X} (t) = \rho_{X}(\eta(t))$. Clearly, there exists a number $0<\epsilon < 1$ for which $\eta_{X}(0,\epsilon) \subset \S_{-}$ and $\eta_{X}(1-\epsilon,1) \subset \S_{+}$. Therefore, since $\Sigma$ separates $\S$, there exists $t_0 \in [\epsilon, 1-\epsilon] $ for which $\eta_{X}(t_0) \in \Sigma$. But then $\eta_X(t_0) \in \Sigma \cap I^{-}(p) \neq \emptyset$, and then $I^{+}(\Sigma) \cap \partial I^{+}(\Sigma) \neq \emptyset$, contradicting the fact that $I^{+}(\Sigma)$ is open. Thus the claim is established. 

From standard results on the structure of achronal boundaries, ${\cal H}_{\pm}$ are $C^0$ achronal connected hypersurfaces in $M$.  

Write $T := {\cal H}_{+} \cup \Sigma$. We now claim that he restriction $\rho_X : {\cal H}_{+} \rightarrow \S$ maps ${\cal H}_{+}$ into a connected open subset of $\S_{+}$ (and indeed it is a homeomorphism onto its image by Invariance of Domain). Otherwise, if $\rho_X(p) \in \S_{-} \cup \Sigma$ for some $p\in {\cal H}_{+}$, then we could pick a future-directed timelike curve from  $\rho_X(p)$ to $p$, compose it with an outward-pointing null curve from $p$ to $\Sigma$, and project it through $\rho_X$ onto $\S$, and a repetition of the above argument for $\eta$ would yield a contradiction. 

We also claim that that $\rho_X(T) = \overline{\S}_{+}$. Indeed, clearly $\rho_{X}(T)\subseteq \overline{\S}_{+}$, and $\Sigma \subseteq T$, so $\Sigma = \rho_{X}(\Sigma) \subseteq \rho_{X}(T)$. Thus $\overline{\S}_{+} \setminus \rho_{X}(T) \subseteq \S_{+}$. Suppose that the claim is false. Then $\overline{\S}_{+} \setminus \rho_{X}(T) \neq \emptyset$, and we must have $\partial_{\S}\rho_{X}(T) \cap \S_{+} \neq \emptyset$. Hence, pick $p \in \partial_{\S}\rho_{X}(T) \cap \S_{+}$. Now, $\rho_{X}(T)$ is clearly closed, so $p = \rho_{X}(q)$ for some $q \in T$. If $q\in \Sigma$, then $\rho_{X}(q)=q=p \in \S_{+}$, a contradiction. Therefore, we can assume that $ q \in {\cal H}_{+}$. Since the latter set is a topological hypersurface, and the restriction of $\rho_X$ to it is a homeomorphism onto its image, we can choose a neighbourhood ${\cal V}_0$ of $q$ in $M$ with ${\cal V}_0 \cap \S = \emptyset$, ${\cal V}_0 \cap T$ open in  ${\cal H}_{+}$ and $ p \in \rho_{X}({\cal V}_0 \cap T) \subseteq \Sigma_{+}$. Since $\rho_{X}({\cal V}_0 \cap T)$ is open in $\S$ because $\rho_{X}$ is an open map, we conclude that $p$ is in the $\S$-interior of $\rho_{X}(T)$, in contradiction with the fact that $p$ must be in the $\S$-boundary of $\rho_{X}(T)$. Therefore, $\rho_X(T) = \overline{\S}_{+}$ as claimed. 

But now our assumption about $\overline{\S}_{+}$ establishes that $T$ is not compact. 

For the remaining part of the proof, fix a complete Riemannian metric $h$ on $M$ with distance function $d_h$.

Now, let $(q_n)$ be any sequence in ${\cal H}_{+}$. For each $n \in \mathbb{N}$, there exists a future-directed, future-inextendible causal curve $\gamma_n:[0, +\infty) \rightarrow M$ parametrized by $h$-arc length, such that $\gamma_n(0) \in \Sigma$ and $\gamma_n(t_n) = q_n$, and $\gamma_n(0,t_n] \subset {\cal H}_{+}$ for some $t_n \in (0, +\infty)$. 

Since $\Sigma$ is compact, passing to a subsequence if necessary we can assume that $\gamma_n(0) \rightarrow p \in \Sigma$. By the Limit Curve Lemma, we can assume that there exists a future-directed, future-inextendible causal curve $\gamma:[0,+\infty) \rightarrow M$ with $\gamma(0) =p$ and such that $\gamma_n|_{C} \rightarrow \gamma|_{C}$ $h$-uniformly in compact subsets. 

Since is $T$ not compact, by the Hopf-Rinow theorem in $(M,h)$, it is either not closed, or not bounded in $d_h$. It is obviously closed, so ${\cal H}_{+}$ is not bounded. In that case, the sequence $(q_n)$ could be chosen as diverging to infinity \footnote{We say that a sequence $(p_n)_{n \in \mathbb{N}}$ in $M$ {\em diverges to infinity} if given any compact subset $C \subseteq M$, only finitely many elements of the sequence are contained in $C$ (conf. \cite{BE}, ch. 8).}; then, the sequence $(t_n)$ is not bounded above, and we can assume that $t_n \rightarrow +\infty$. 

Fix a number $t >0$. Eventually, $t_n > t$, in which case $\gamma_n(t) \in \in E^{+}(\Sigma)$. We cannot have $\gamma(t) \in I^{+}(\Sigma)$, for otherwise $\gamma_n (t) \in I^{+}(\Sigma)$ for large $n$ and therefore $q_n \in I^{+}(\Sigma)$, a contradiction. Therefore $\gamma(t) \in E^{+} (\Sigma)$, and since $\gamma(t) \notin \Sigma$ from the acausality of $\S$, we conclude that $\gamma \subset T$. Since the latter set is achronal, $\gamma$ must be a reparametrization of a future-directed, outward-pointing null $\Sigma$-ray normal to $\Sigma$ as desired (see Theorem 51, p. 298 of Ref. \cite{oneill}). 

\qcd

\begin{remark}
\label{remark1}
The conclusion of Proposition \ref{mainproposition1} is false without the piercing assumption. {\em Indeed, the following simple example shows this. Consider the flat 2D Lorentzian cylinder $S^1 \times \mathbb{R}$ obtained by isometrically identifying 2D Minkowski spacetime along the spatial direction. Fix any acausal circle $C$ going around the cylinder with unit normal parallel to the ``axis'' of the cylinder, and fix two antipodal points $p$ and $q$ on $C$. Now, delete the whole causal past of $q$ (including $q$ itself). We denote the resulting manifold by $N_0$ and its (flat) metric by $g_0$. Choose any compact Riemannian manifold $(N,h)$. Let $M = N_0 \times N$, $g = g_0 \oplus h$, $\S = (C \setminus \{q\}) \times N$ and $\Sigma = p \times N$. This spacetime is globally hyperbolic, and $\S$ is a partial Cauchy hypersurface (but not Cauchy). Clearly, there are no future-directed null $\Sigma$-rays emanating from $\Sigma$.} 
\end{remark}    

The next proposition uses the following result, presented here as a lemma, and proven in Ref. \cite{me2}.

\begin{lemma}
\label{lemma1}
Let $\Sigma \subset M$ be a surface contained in a partial Cauchy hypersurface $\S$. Then, at least one of the following alternatives occurs: 
\begin{itemize}
\item[i)] $\Sigma$ is a future-trapped set, 
\item[ii)] there exists an inward-pointing $\Sigma$-ray starting at $\Sigma$, 
\item[iii)] there exists an outward-pointing $\Sigma$-ray starting at $\Sigma$. 
\end{itemize}
\end{lemma} 

\qcd

\begin{remark}
\label{remark2}
{\em This result cannot be improved without further assumptions. To see this one can consider the spacetime constructed in Remark \ref{remark1}. $\Sigma$ is a future-trapped set therein, but one can easily delete suitable points along the generators of $\partial I^{+}(\Sigma)$ to show that it can be have an inward-pointing $\Sigma$-ray but not an outward-pointing one or vice-versa. But there is an additional assumption which can be quite natural in examples. This is that $\Sigma$ {\em bounds} (in $\S$), i.e, that $\overline{\S}_{-}$ is compact. This will be a key assumption in the next result.}
\end{remark}

\begin{proposition}
\label{mainproposition2}
Let $\Sigma \subset M$ be a surface contained in a partial Cauchy hypersurface $\S$. Suppose that $\Sigma$ bounds. Then, either $\Sigma$ is a future-trapped set, or else there exists an outward-pointing $\Sigma$-ray starting at $\Sigma$. 
\end{proposition} 

{\em Proof.} Suppose that $\Sigma$ is not a future-trapped set. We must show that it does not admit a inward-pointing $\Sigma$-ray, and so by Lemma \ref{lemma1}, there exists an outward-pointing $\Sigma$-ray starting at $\Sigma$. 

Suppose first that $(M,g)$ is globally hyperbolic and that $\S$ is a Cauchy hypersurface. In that case, using the notation in the proof of Proposition \ref{mainproposition1}, and by completely analogous arguments using the piercing, we can show that ${\cal H}_{-} \cup \Sigma$ must be compact, as it is homeomorphic to $\overline{\S}_{-}$. But any inward-pointing null $\Sigma$-ray would have to be contained in ${\cal H}_{-} \cup \Sigma$, which cannot occur since the spacetime is in particular strongly causal and because the latter set is compact. Thus such a ray cannot exist. 

Now, let us turn to the general case. Since $\S$ is a partial Cauchy hypersurface, its Cauchy development (domain of dependence) $D(\S)$ is open, and viewed as a spacetime in its own right, it is globally hyperbolic with $\S$ as a Cauchy hypersurface (see, e.g., \cite{oneill} ch. 14). Suppose there exists an inward-pointing $\Sigma$-ray $\eta:[0,a) \rightarrow M$ starting at $\Sigma$. Then it is contained in $\partial I^{+} (\Sigma)$. However, it is easy to check that, due to the fact that $D(S)$ is open and causally convex (in the sense that causal curves cannot leave and reenter it), that $ \partial_{D(\S)} I^{+} (\Sigma,D(\S)) = \partial I^{+} (\Sigma) \cap D(\S)$. Thus, the portion of $\eta$ contained in $D(S)$ would be an inward-pointing $\Sigma$-ray in $(D(\S),g|_{D(\S)})$, contradicting the conclusion in the previous paragraph. This last contradiction completes the proof.

\qcd

\section{Main Results \& Consequences}\label{mainstuff}


We are now ready to state and prove our main results. The first one is a Hawking-Penrose-type theorem \cite{penrose,BE} in the presence of a generic MOTS that bounds. Actually, the proof also yields a singularity theorem for {\em outer trapped surfaces} (which bound), and we include this case as well. Using the notation in the previous Section, we say that $\Sigma$ is {\em outer trapped} if $\theta_{+} <0$.  
 
\begin{theorem}
\label{mainsingularity1}
Let $(M,g)$ be a spacetime of dimension $n \geq 3$ satisfying the the following requirements:
\begin{itemize}
\item[(1)] $(M,g)$ contains no closed timelike curve.
\item[(2)] $(M,g)$ obeys the timelike convergence condition, i.e. its Ricci tensor satisfies $Ric(v,v)\geq 0$, for all timelike vectors $v$.
\item[(3)] $(M,g)$ contains a surface $\Sigma$ contained in a partial Cauchy hypersurface which bounds, and which is either a generic MOTS or else it is outer trapped.
\item[(4)] The nonspacelike generic condition holds in $(M,g)$ \footnote{Recall that the nonspacelike generic condition holds in $(M,g)$ when there are non-zero tidal accelerations $\langle v,R(v,\gamma ')\gamma' \rangle $ along {\em any} inextendible nonspacelike geodesic $\gamma: (a,b) \rightarrow M$ (see, e.g. Chs. 2, 12 and 14 of \cite{BE} for a thorough discussion of the generic condition).}
\end{itemize}
Then $(M,g)$ is nonspacelike geodesically incomplete
\end{theorem}

{\em Proof.} Suppose first that $\Sigma$ is not a future-trapped set. Then, by Prop. \ref{mainproposition2} there exists a future-directed, affinely parametrized, outward-pointing, null $\Sigma$-ray $\eta:[0, a) \rightarrow M$ normal to $\Sigma$. Suppose $\eta$ is future-complete, so that we put $a= +\infty$. Since $\eta$ is a $\Sigma$-ray, a standard analysis (see Ch. 12 of \cite{BE}) there exists a globally defined Lagrange tensor field $A(t)$ on $\eta$ such that $b(t) = A'(t)A^{-1}(t)$ obeys the Riccati-type equation
\begin{equation}
\label{riccati}
b' + b^2 + {\cal R} =0,
\end{equation}
where ${\cal R}(t) : N(\eta(t)) \rightarrow N(\eta(t))$ takes the values of $R(v,\eta')\eta'$ on normal vectors $v \bot \eta(t)$ (taken modulo $\eta'$), $N(\eta(t))$ being their span, and so that $\theta_b(t) \equiv tr(b(t))$ satisfies $\theta_b(0) = \theta_{+}(\eta(0)) \leq 0$. Now, tracing Eq. (\ref{riccati}) we obtain the Raychaudhuri equation
\begin{equation}
\label{raychau}
\theta'_b + \frac{1}{n-2} \theta^2_b = - Ric(\eta',\eta') - \sigma ^2,
\end{equation}    
where the shear scalar $\sigma$ is the trace of the square of the trace-free part of $b$. If we define $f:[0,+\infty) \rightarrow \mathbb{R}$ by
\begin{equation}
\label{convenience}
f(t) = - Ric(\eta'(t),\eta'(t)) - \sigma ^2(t),
\end{equation}
the assumption on the Ricci tensor implies that $f\leq 0$, so that integrating Eq. (\ref{raychau}) from $0$ up to an arbitrary $s$, we get
\[
\limsup_{s \rightarrow + \infty} \left(\theta_b(s) + \frac{1}{n-2}\int^{s}_0 \left(\theta_b(t)\right)^2 dt \right) = \theta_b(0) + \limsup_{s \rightarrow + \infty} \int^s_0 (f)dt \geq 0,
\]
where we have used a result proven by Guimaraes \cite{guima} in the last inequality (see also \cite{me2}). We conclude that $\theta_b \equiv 0$, and again by Eq. (\ref{raychau}), that $f \equiv 0$. In particular, we must have $\theta_{+}(\eta(0)) =0$, which already yields a contradiction in the case when $\Sigma$ is outer trapped. But also, $b \equiv 0$ along $\eta$ and hence ${\cal R} \equiv 0$ by Eq. (\ref{riccati}), contradicting the generic condition on normal geodesics. 

Therefore, either $\eta$ is future incomplete and we are done, or $\Sigma$ is a future-trapped set. In the latter case, the standard Hawking-Penrose singularity theorem \cite{penrose,HE,BE} now yields the conclusion, so the proof is complete. 

\qcd

\begin{remark}
{\em As mentioned in the introduction, in spacetime dimension $n=4$ a celebrated result by Schoen and Yau establishes that compact MOTS appear naturally for large concentrations of matter \cite{SY3}. In their setting, $\S$ is a three-dimensional and asymptotically flat manifold bearing initial data. In the particular case when $\S$ is diffeomorphic to $\mathbb{R}^3$, then any closed embedded MOTS will bound \footnote{I thank Greg Galloway for suggesting this example, together with Ref. \cite{SY3}.}.}
\end{remark} 

In Section 3 of Ref. \cite{EGP}, the authors call an {\em initial data singularity theorem} any result which proves the existence of closed trapped surfaces in initial data sets from suitable conditions on the geometry and/or energy-momentum density. Of course, what is really relevant for most singularity theorems are trapped {\em sets}, closed trapped surfaces being just one class of such sets when some additional conditions hold (strong causality plus the null convergence condition suffice in this case). However, they are particularly important for, say, numerical Relativity, because the signs of the expansion scalars can be ascertained in a compact region of the initial data underlying manifold, and can be expressed solely in terms of the initial data. MTSs or MOTS also are likewise quasi-locally detectable. In particular, the proof of Theorem \ref{mainsingularity1} actually means that any proof of the existence of a bounding MOTS in initial data sets can also be viewed as an initial data singularity theorem in this sense:   

\begin{theorem}
\label{mainsingularity3}
Let $(M,g)$ be a spacetime of dimension $n \geq 3$ in which the null convergence condition holds (i.e., $Ric(v,v) \geq 0$ for all null vectors $v \in TM$), and which contains a generic MOTS or outer trapped $\Sigma$ in a partial Cauchy hypersurface. Then, either $(M,g)$ is null geodesically incomplete, or $\Sigma$ is a future-trapped set. 
\end{theorem}

There is a special setting which is also of interest, that of {\em strictly stable} MOTS. It is well-known that MOTS have a notion of stability \cite{AEM,AMMS,AM} similar to that of minimal surfaces. Let us briefly recall the setting and refer the reader to Refs. \cite{AEM,AMMS,AM,mots2} and references therein for further details and proofs of the facts mentioned. Consider  a {\em normal variation}, i.e., a variation $t \rightarrow \Sigma_t$ in $\S$ with $\Sigma_0 = \Sigma$ and variation vector field $V = \phi N_{+}$, with $\phi \in C^{\infty}(\Sigma)$. Then one can show that
\begin{equation}
\label{variation}
\frac{\partial \theta_{+}}{\partial t}| _{t=0} = L\phi,
\end{equation}
where $L: C^{\infty}(\Sigma) \rightarrow C^{\infty}(\Sigma)$ is an elliptic linear operator, the {\em stability operator}, whose specific form need not concern us here. Suffice it to say that $L$ has an eigenvalue with smallest real part $\lambda_0$ which is real and admits a strictly positive smooth eigenfunction $\phi_0$. Now, the MOTS $\Sigma$ is said to be {\em stable} [resp. {\em strictly stable}] if $\lambda_0 \geq 0$ [resp. $\lambda_0 >0$. (An important situation in which a MOTS is stable \cite{mots2} is when it is {\em outermost}, i.e., when there are neither outer trapped surfaces nor MOTS outside of $\Sigma$ which are homologous to $\Sigma$.)
  
Therefore, if $\Sigma$ is strictly stable, we can use $\phi_0$ in our variation, and in that case Eq. (\ref{variation}) implies that for some $\epsilon >0$, $\Sigma_{- \epsilon}$ is an outer trapped closed surface. It can be chosen to bound if $\Sigma$ bounds. Taking these facts into consideration yields the following.  
 
\begin{corollary}
\label{maincorolary}
Let $(M,g)$ be a spacetime of dimension $n \geq 3$ satisfying the the following requirements:
\begin{itemize}
\item[(1)] $(M,g)$ contains no closed timelike curve.
\item[(2)] $(M,g)$ obeys the timelike convergence condition, i.e. its Ricci tensor satisfies $Ric(v,v)\geq 0$, for all timelike vectors $v$.
\item[(3)] $(M,g)$ contains a strictly stable MOTS $\Sigma$ contained in a partial Cauchy hypersurface and which bounds therein.
\item[(4)] The nonspacelike generic condition holds in $(M,g)$.
\end{itemize}
Then $(M,g)$ is nonspacelike geodesically incomplete
\end{corollary}
\qcd

If we do not assume that the MOTS bounds, then we still obtain a singularity theorem provided we strengthen the causal assumption on $(M,g)$ by using Proposition \ref{mainproposition2}: 

\begin{theorem}
\label{mainsingularity2}
Let $\Sigma \subset M$ be a surface contained in a partial Cauchy hypersurface $\S$. Suppose that the following holds.
\begin{itemize}
\item[i)] $(M,g)$ is causally simple and obeys the null convergence condition. 
\item[ii)] $\overline{\S}_{+}$ is noncompact and $\S$ admits a piercing $X$. 
\item[iii)] $\Sigma$ is either a generic MOTS or else it is outer trapped.
\end{itemize}
Then $(M,g)$ is null geodesically incomplete. 
\end{theorem} 

{\em Proof.} By Prop. \ref{mainproposition1}, conditions (i) and (ii) imply that there exists a future-directed, affinely parametrized, outward-pointing, null $\Sigma$-ray $\eta:[0, a) \rightarrow M$ normal to $\Sigma$. A repetition of the proof of Theorem \ref{mainsingularity1} (without the trapped case, which does not apply here) now yields the conclusion. 

\qcd

\begin{acknowledgements}
I wish to thank Greg Galloway for valuable correspondence on the subject of this paper. 
\end{acknowledgements}


%
\end{document}